\documentclass[pre,aps,floats,superscriptaddress,floatfix,twocolumn]{revtex4}
\usepackage{amssymb,amsmath}
\usepackage{amsmath,amssymb}
\usepackage{graphicx}
\usepackage{psfrag}
\usepackage{color}
\usepackage{dcolumn}
\usepackage{bm}
\usepackage[normalem]{ulem}
\usepackage{gensymb}

\def\beq{\begin{equation}}
\def\eeq{\end{equation}}
\def\bea{\begin{eqnarray}}
\def\eea{\end{eqnarray}}
\begin{document}
\title{Conserved Kardar-Parisi-Zhang equation: Role of quenched disorder in determining universality}
\author{Sudip Mukherjee}\email{sudip.bat@gmail.com, sudip.mukherjee@saha.ac.in}
\affiliation{Barasat Government College,
10, KNC Road, Gupta Colony, Barasat, Kolkata 700124,
West Bengal, India}
\affiliation{Condensed Matter Physics Division, Saha Institute of
Nuclear Physics, Calcutta 700064, West Bengal, India}

\date{\today}

\begin{abstract}
 We study the stochastically driven conserved Kardar-Parisi-Zhang (CKPZ) equation with quenched disorders. Short-ranged quenched disorders is found to be a relevant perturbation on the pure CKPZ equation at one dimension, and as a result, a new universality class different from pure CKPZ equation appears to emerge. At higher dimensions, quenched disorder turns out to be ineffective to influence the universal scaling. This results in the asymptotic long wavelength scaling to be given by the linear theory, a scenario identical with the pure CKPZ equation. For sufficiently long-ranged quenched disorders, the universal scaling is impacted by the quenched disorder even at higher dimensions. 
\end{abstract}

\maketitle

\section{Introduction}

The idea of universality,
parametrised by the space dimensions, symmetries and the order
parameter components, allows one to have a systematic physical understanding of universal 
scaling properties near the critical points and in the broken symmetry phases in equilibrium systems~\cite{fisher,chaikin}. Subsequently, the concept of universality classes have been extended to systems out of equilibrium. A notable example of a nonequilibrium universality class is the Kardar-Parisi-Zhang (KPZ) universality class~\cite{kardar,natter,stanley}. Nonequilibrium universalities are still topics of intense research in theoretical physics.

Conservation laws are known to be important in physics. In equilibrium systems, the presence or absence of conservation laws only affect the dynamic universality, i.e., the relaxation of the fluctuations and the time-dependent correlation functions, while the time-independent or the thermodynamic properties remain unaffected by it. For instance, in pure relaxational dynamic of the Ising model near its critical point, conservation of the magnetisation leads to a slower relaxation of the fluctuations than when it is not~\cite{chaikin,halpin}. In contrast, in nonequilibrium systems conservation laws affect even the time-independent quantities. A classic example of this is the conserved KPZ (CKPZ) equation, which like the KPZ equation describes a fluctuating surface, but now with a conservation law~\cite{ckpz}. The CKPZ equation shows distinctly different scaling properties of the equal-time correlation function of the height fluctuations, and in so far as even the time-independent properties are concerned, necessarily belongs to a universality class different from the original KPZ universality~\cite{ckpz}. For instance, the roughness and the dynamic exponents of the height fluctuations in the KPZ equation obey an exact relation that arises due to the Galilean invariance of the KPZ equation.  In contrast, the CKPZ equation is {\em not} Galilean invariant~\cite{ckpz-janssen}, and as a result, there is no corresponding exact relation between the scaling exponents in the CKPZ equation. In addition the KPZ equation undergoes a roughening transition between a smooth and a perturbatively-inaccessible rough phase~\cite{kardar} at dimension $d>2$. This has no analogue for the pure CKPZ equation.

Quenched disorder are known to affect the macroscopic properties of equilibrium systems. For example, even arbitrarily weak random
fields are known to destroy long-ranged ferromagnetic order in all spatial dimensions $d\leq 4$~\cite{rand-fld}. Similarly, quenched disorders that introduces local quenched fluctuations in the critical temperatures introduces new universality classes different from the corresponding pure model~\cite{rand-tc,sudip}. { Effects of quenched disorder on nonequilibrium systems are expected to more dramatic, given the sensitive dependences of nonequilibrium universality classes on all kinds of perturbations. In the absence of any general framework to study nonequilibrium systems, it is useful to construct and study simple nonequilibrium models with quenched disorders that are amenable to analytical studies and hence allow for systematic enumeration of physical quantities within simple calculational set ups. Such a study should be useful in forming general understanding of the effects of quenched disorders on nonequilibrium universality classes.}

In this article, we study a version of quenched disordered CKPZ equation. We evaluate the universal scaling properties, and compare and contrast them with the corresponding results for the pure CKPZ equation. We consider both short-ranged and long-ranged quenched disorders. We show that for short-ranged quenched disorders, the universal scaling properties are affected by the disorder at one dimension ($1d$), leading to a new universality class, whereas at dimensions two or more, quenched disorder is irrelevant. For long-ranged quenched disorder, the dimension at which quenched disorder ceases to be relevant is higher than two, and can in fact be varied by tuning the spatial scaling of the variance of the long-ranged disorder. The rest of the article is organised as follows. In Sec.~\ref{model}, we set up the CKPZ equation with quenched disorders. Next we discuss the scaling in the linearised limit in Sec.~\ref{lin-scal}. We then set up a dynamic RG calculation in Sec.~\ref{rg-an}. We separately calculate the scaling exponents for short-ranged and long-ranged disorder in Sec.~\ref{short-dis} and Sec.~\ref{long-dis}, respectively. We summarise our results in Sec.~\ref{summ}. We have used one-loop dynamic renormalisation group (RG) calculations for our work. We provide the necessary technical details in Appendix for interested readers.

\section{Conserved KPZ equation with quenched disorder} \label{model}

We generalise the CKPZ equation~\cite{ckpz} in the presence of quenched disorder. The precise form of the quenched disordered conserved dynamical equation for the height field $h$ should depend on {\em how exactly} the quenched disorder coupled with $h$. To make the ensuing study concrete, we consider quenched disorder given by a quenched vector field $F_i({\bf r})$, where $i$ refers to the Cartesian component and $\bf r$ is the position vector. Further, we assume that ${\bf F}$ couples with the spatial nonuniformities of the height field, i.e., with ${\boldsymbol\nabla} h$ to the leading order in spatial gradients.  In order to reduce the number of model parameters and simplify the situation, we further impose the condition that in the absence of any time-dependent of annealed noise if $h=const.$ at some time $t$, it remains so at all other times, just as the pure CKPZ equation. Secondly noting that the conserved current ${\bf J}_{CKPZ}$ in the pure CKPZ equation has the form~\cite{ckpz}
\begin{equation}
 {\bf J}_{CKPZ}({\bf r},t)=-{\boldsymbol\nabla} \left[\nu\nabla^2 h + \frac{\lambda}{2} ({\boldsymbol\nabla}h)^2\right],
\end{equation}
one has ${\bf J}_{CKPZ} ({\bf k}=0,t)=0$, where $\bf k$ is a Fourier wavevector; ${\bf J}_{CKPZ}({\bf k},t)$ is the spatial Fourier transform of ${\bf J}_{CKPZ}({\bf r},t)$. 
We then generalise ${\bf J}_{CKPZ}$ in the presence of quenched disorder. We write the corresponding conserved current $\bf J$ for the disordered CKPZ equation as
\begin{equation}
 {\bf J}({\bf r},t)=-{\boldsymbol\nabla} \left[\nu\nabla^2 h + \frac{\lambda}{2} ({\boldsymbol\nabla}h)^2 + \lambda_1 {\bf F}\cdot {\boldsymbol\nabla} h\right].\label{dis-curr}
\end{equation}
Thus, $ {\bf J}({\bf k}=0,t) =0$, just like the vanishing of ${\bf J}_{CKPZ}({\bf k}=0)$; ${\bf J}({\bf k},t)$ is the spatial Fourier transform of ${\bf J}({\bf r},t)$.
With this proviso, we write down the minimal quenched disordered CKPZ equation in the long wavelength limit:
We have 
\begin{equation}
 \frac{\partial h}{\partial t}= - {\boldsymbol\nabla}\cdot {\bf J}
\end{equation}
giving
\begin{equation}
 \frac{\partial h}{\partial t} = -\nabla^2 \left[\nu\nabla^2 h + \frac{\lambda}{2} ({\boldsymbol\nabla}h)^2 + \lambda_1 {\bf F}\cdot {\boldsymbol\nabla} h\right] + \eta.\label{ckpz-dis}
\end{equation}
Here, $\nu>0$ is a damping coefficient and $\lambda$ and $\lambda_1$ are nonlinear coupling constants both of which can be any sign. Lastly, the annealed or time-dependent noise $\eta({\bf r},t)$ is assumed to be zero-mean Gaussian-distributed with a variance
\begin{equation}
 \langle \eta({\bf r},t)\eta(0,0)\rangle = -2D_h \nabla^2\delta({\bf r})\delta (t),\label{anneal-noise}
\end{equation}
that is consistent with the conservation law form for the dynamics of $h$. Evidently, independent of the specific form of the quenched disorder, $h=const.$ satisfies Eq.~(\ref{ckpz-dis}) at all times $t$ so long as $\eta=0$, in exact analogy with the pure CKPZ equation. Equation~(\ref{ckpz-dis}) upon setting $\lambda_1=0$ evidently reduces to the well-known pure CKPZ equation~\cite{ckpz}. 
In order to completely describe the model, the distribution of $F_i$ must be provided. Quenched disorder $F_i$ is assumed to be zero-mean Gaussian distributed with a given variance. We consider both short- and long-ranged quenched disorder. For short-ranged quenched  disorder, the variance reads
\begin{equation}
 \langle F_i({\bf r})F_j(0)\rangle = 2D_F \delta_{ij}\delta({\bf r}),\label{short-vari}
\end{equation}
where as for the long-ranged case we choose
\begin{equation}
 \langle F_i({\bf r})F_j(0)\rangle = 2D_F \delta_{ij}|r|^{\alpha-d},\label{long-vari}
\end{equation}
where $\alpha>0$ parametrises the disorder distribution, i.e., it describes {\em ``how long''} is  the long-ranged disorder; further amplitude $D_F>0$.   We choose $0<\alpha<d$, implying that the disorder correlation has a range longer than $\delta({\bf r})$, which nonetheless decays as the separation $|r|$ increases. It is convenient to express variances (\ref{short-vari}) and (\ref{long-vari}) in the Fourier space. For the short-ranged case we get
\begin{equation}
 \langle F_i({\bf k},\omega)F_j({\bf k'},\omega')\rangle = 2D_F \delta_{ij}\delta ({\bf k+k'})\delta(\omega+\omega')\delta(\omega).\label{short-vari-k}
\end{equation}
Likewise, the variance in the long-ranged case is
\begin{equation}
 \langle F_i({\bf k},\omega)F_j({\bf k'},\omega')\rangle = 2D_F\delta_{ij}|k|^{-\alpha}\delta({\bf k+k'}) \delta(\omega+\omega')\delta(\omega).\label{long-vari-k}
\end{equation}
Here, $\bf k,k'$ are Fourier wavevectors, and $\omega,\omega'$ are Fourier frequencies. The factor of $\delta(\omega)$ that appears in both (\ref{short-vari-k}) or (\ref{long-vari-k}) has its origin in the fact that $F_i({\bf r})$ is time independent. Equation~(\ref{ckpz-dis}) together with the variances (\ref{short-vari}) or (\ref{long-vari}), along with the annealed noise (\ref{anneal-noise}) completely define the model. 

{ Before we embark upon calculating the scaling exponents for (\ref{ckpz-dis}), it is instructive to discuss the physical implication of the disorder $\lambda_1$-term in (\ref{ckpz-dis}) or (\ref{dis-curr}). Current ${\bf J}$ in (\ref{dis-curr}) may be written as ${\bf J}=-{\boldsymbol\nabla}\mu$, where the disorder-dependent local {\em chemical potential} $\mu$ is
\begin{equation}
 \mu\equiv \nu\nabla^2 h + \frac{\lambda}{2} ({\boldsymbol\nabla}h)^2 + \lambda_1 {\bf F}\cdot {\boldsymbol\nabla} h,
\end{equation}
where the last term $\lambda_1 {\bf F}\cdot {\boldsymbol\nabla} h$ is the disorder contribution to $\mu$. Thus the quenched disordered CKPZ equation (\ref{ckpz-dis}) may be interpreted as modeling local height fluctuations in surface diffusion in the presence of quenched disorder. { In a microscopic realisation of this process, this quenched disorder may arise, e.g., in the local deformations or heterogeneities affecting the diffusion, of any underlying lattice, on which the surface diffusion 
may take place. Such quenched inhomogeneities may locally facilitate or hinder pure surface diffusion, depending upon the microscopic forms of the disorder locally. In appropriate experimental realisations, these results may be tested by measuring surface diffusion on disordered substrates (e.g., in molecular beam epitaxy experiments with quenched disorder).} }

\section{Scaling}\label{scaling}


 We are interested in the scaling of the correlation function
\begin{equation}
 C( r,t)\equiv \langle h({\bf r},t)h(0,0)\rangle=|r|^{2\chi}f(|r|^z/t),\label{creal}
\end{equation}
or its Fourier transformed version
\begin{equation}
 C(k,\omega)\equiv \langle |h({\bf k},\omega)|^2\rangle=k^{2\chi_k}\tilde f(k^z/\omega)\label{cfour}
\end{equation}
in the long wavelength limit. Here $\chi$ and $z$ are the roughness and dynamic exponents, respectively; $\chi_k$ can be connected to $\chi$ by Fourier transform, giving
\begin{equation}
 \chi_k=-d-\chi-z.
\end{equation}
Further, $f(|r|^z/t)$ and $\tilde f(k^z/\omega)$ are dimensionless scaling function of their respective arguments.

\subsection{ Linear theory}\label{lin-scal}

{ The linear limit of Eq.~(\ref{ckpz-dis}) is obtained by dropping all nonlinear terms (in this case quadratic in ${\boldsymbol\nabla}h$ or bilinear in ${\bf F}$ and ${\boldsymbol\nabla} h$); see Eqs.~(\ref{action1}) and (\ref{action2})}. Obviously, 
in the linear limit, both the pure CKPZ equation and our model equation are identical  and the correlation function is known {\em exactly}. We have
\begin{equation}
 C(k,\omega)=\frac{2D_h k^2}{\omega^2 + \nu^2k^8}.\label{lin-corr}
\end{equation}
This implies the {\em exact values} $z=4$ and $\chi=1-d/2$. It remains to be seen
how the various nonlinear terms may affect these scaling
exponents in the linear theory. 

\subsection{Anharmonic effects}\label{rg-an}

Presence of the nonlinear terms no longer allows enumeration of the exact scaling exponents for (\ref{ckpz-dis}). Unlike in the linear theory, exact enumeration of the scaling exponents is no longer possible due to the nonlinear terms. Thus perturbative treatments are necessary. Similar to the pure CKPZ equation~\cite{ckpz}, na\"ive perturbation theory produces diverging corrections to the model parameters. These divergences may be systematically handled within the framework of dynamic RG~\cite{halpin}. 

Although the dynamic RG procedure by now is well-documented~\cite{halpin} in the standard literature, we give below a brief
outline of the method for the convenience of the readers. It is useful to first cast the dynamical equation (\ref{ckpz-dis}) into a dynamic generating functional by introducing a conjugate field $\hat h({\bf r},t)$; see Ref.~\cite{janssen}, see also Appendix~\ref{short-action} for some details. The dynamic generating functional is then averaged over the Gaussian disorder distribution with variances (\ref{short-vari}) or (\ref{long-vari}). The momentum shell dynamic RG procedure consists of integrating over the short wavelength Fourier modes of $h({\bf r},t)$ and $\hat h({\bf r},t)$ in the generating functional. This is then followed by  rescaling of lengths
and time. In particular, we follow the standard approach of initially restricting wavevectors to lie in a Brillouin zone: $|q| < \Lambda$, where $\Lambda$ is an ultra-violet cutoff,
which should be of order the inverse of the lattice spacing $a$,
although its precise value is unimportant so far as the scaling in the long wavelength limit is considered. The height field $h({\bf r},t)$ and its dynamic conjugate $\hat h ({\bf r},t)$ are then split into two parts, a high and low wave vector parts $h({\bf r},t)=h^>({\bf r},t) + h^<({\bf r},t)$ and $\hat h({\bf r},t)=\hat h^>({\bf r},t) + \hat h^<({\bf r},t)$, where $h^>({\bf r},t)$ and $\hat h^>({\bf r},t)$ are non-zero in the high wavevector range $\Lambda/b< k < \Lambda,\,b>1$, whereas $h^<({\bf r},t)$ and $\hat h^<({\bf r},t)$ are non-zero in the low wavevector range $k< \Lambda/b$. Next, $h^>({\bf r},t)$ and $\hat h^> ({\bf r},t)$ are to be integrated out in the dynamic generating functional. Of course, this integration cannot be done exactly, but is done
perturbatively in the anharmonic couplings in (\ref{action1}) for short-ranged disorder or (\ref{action2}) for long-ranged disorder. 
This perturbation theory is usually represented
by Feynman diagrams, with the order of perturbation
theory given by the number of loops in the diagrams that
we calculate; see Appendix~\ref{short-appen} and Appendix~\ref{long-diag}. Next to this perturbative step, we rescale length by ${\bf r}={\bf r'}b$,
in order to restore the UV cutoff back to $\Lambda$. We further rescale time by $t = t'b^z$, where $z$ is the
dynamic exponent. This is then followed by rescaling of $h^<({\bf r},t)$ and $\hat h^<({\bf r},t)$,
the long wave length parts of $h({\bf r},t)$ and $\hat h({\bf r},t)$; see Appendix~\ref{resc}.

We separately study this problem with short-ranged and long-ranged disorders. We confine ourselves
to a low-order (one-loop) RG analysis, following the calculational scheme outlined above. 

\subsubsection{Short-ranged disorder}\label{short-dis}

We provide the one-loop Feynman diagrams for the model parameters in Appendix~\ref{short-appen}. The Feynman graphs for the diffusivity $\nu$ and the nonlinear vertices $(\lambda/2)(\nabla^2\hat h) ({\boldsymbol\nabla}h)^2$ and $\lambda_1^2 D_F (\nabla^2 \hat h({\bf r},t_1))\nabla_m h({\bf r},t_1)(\nabla^2 \hat h({\bf r},t_2))\nabla_m h ({\bf r},t_2)$ [see action (\ref{action1})] are shown in Appendix~\ref{short-appen}. 

At the one-loop order diffusivity $\nu$ receives two fluctuation corrections, one of which survives in the pure limit and other one originates from the disorder vertex. The relevant Feynman diagrams are shown in 
Appendix~\ref{short-appen}.

For reasons similar to the pure CKPZ equation, there are {\em no} one-loop fluctuation corrections to the annealed noise strength $D_h$, since these one-loop diagrams are all ${\cal O}(k^4)$, where as the bare or unrenormalised noise strength is ${\cal O}(k^2)$. 

Vertex coefficients $\lambda_1^2 D_F$ and $\lambda$ are themselves renormalised at the one-loop order by the graphs shown in 
Appendix~\ref{short-appen}.

We follow this diagrammatic expansion by rescaling the long wavelength part of the height field $h({\bf r},t)$ as
\begin{equation}
 h({\bf r},t)=\zeta h'({\bf r'},t');\;\zeta = b^\chi,\label{h-scale}
\end{equation}
Exponents $\chi$ and $z$ are to be chosen to produce a fixed point. This procedure together with $b=e^{dl}\approx 1+dl$, ultimately leads to the following recursion relations:
\begin{eqnarray}
 \frac{d\nu}{dl}&=&\nu\left[ z-4 + g\frac{4-d}{4d} + \tilde g \frac{d-2}{d}\right],\label{recur-short1}\\
 \frac{dD_h}{dl}&=&D_h\left[z-2-d-2\chi\right],\label{recur-short2}\\
 \frac{d\lambda}{dl}&=& \lambda\left[z+\chi-4 + 2\tilde g - \frac{4}{d}\tilde g\right],\label{recur-short3}\\
 \frac{d}{dl}\left(\lambda_1^2 D_F\right) &=& \lambda_1^2 D_F \left[2z-d-6+2\tilde g - \frac{6\tilde g}{d}\right].\label{recur-short4}
\end{eqnarray}
Here, $g\equiv \frac{\lambda^2 D_h}{\nu^3}k_d$ and $\tilde g \equiv \frac{\lambda_1^2 D_F}{\nu^2} k_d$ are the two effective dimensionless coupling constants;   $k_d=S_d/(2\pi)^d$, where $S_d$ is the surface area of a hypersphere with unit radius in $d$-dimensions. We now use the flow equations (\ref{recur-short1}-\ref{recur-short4}) to calculate the flow equations for $g$ and $\tilde g$. We obtain
\begin{eqnarray}
 \frac{dg}{dl}&=&g\left[2-d + 4\tilde g - \frac{8}{d}\tilde g - 3g\frac{4-d}{4d}-3\tilde g\frac{d-2}{d}\right],\label{flow1}\\
 \frac{d\tilde g}{dl}&=&\tilde g\left[2-d+2\tilde g-\frac{6\tilde g}{d} - 2g \frac{4-d}{4d} - 2\tilde g\frac{d-2}{d}\right].\label{flow2}
\end{eqnarray}
Before proceeding further, we note the following from the flow equations (\ref{flow1}) and (\ref{flow2}). First of all, setting aside the fluctuation corrections, we note that both scale-dependent $g,\tilde g$ scale as $\exp[(2-d)l]$, where $\exp (l)$ is a length-scale, indicating the ``equal'' relevance of $g(l)$ and $\tilde g(l)$ in a RG sense. Secondly, for $d<(>)2$ both $g(l)$ and $\tilde g(l)$ grow (decay) under rescaling, showing $d=2$ is the critical dimension, same as for the pure CKPZ equation. In fact, flow equations~(\ref{flow1}) and (\ref{flow2}) are amenable to an $\epsilon$-expansion where $\epsilon\equiv 2-d$, and the scaling exponents may be calculated in an $\epsilon$-expansion similar to the pure CKPZ equation~\cite{ckpz}. 

\begin{figure}[htb]
 \includegraphics[width=7cm]{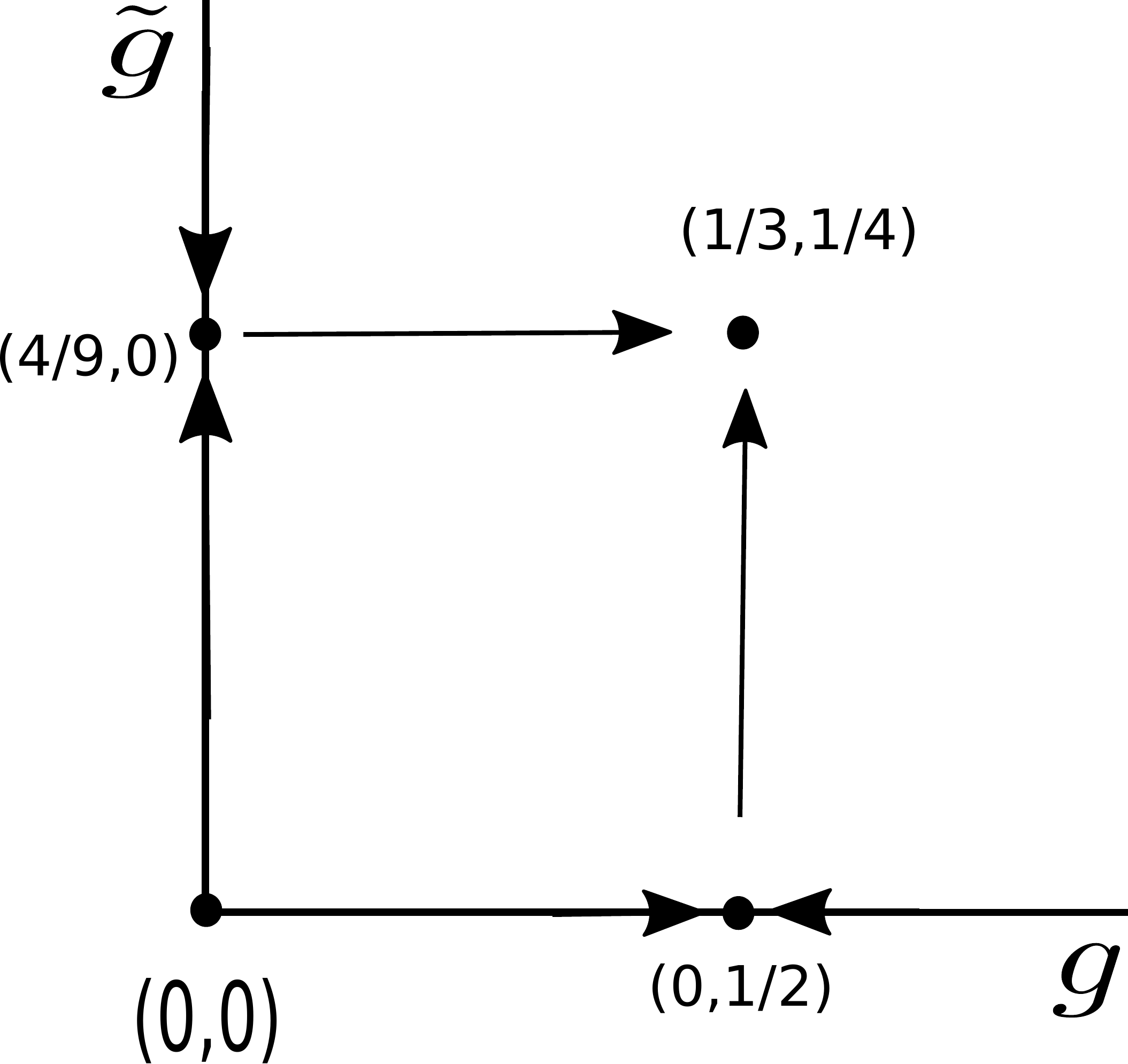}
 \caption{RG flow diagram in the $g-\tilde g$ plane. The fixed point  ($g^*,\,\tilde g^*)=(1/3,1/4)$ is globally stable, fixed points (0,0), (4/9,0) and (0,1/2) are unstable (see text).}\label{flow-short}
   \end{figure}

First consider scaling in one dimension. Instead of using an $\epsilon$-expansion with $\epsilon=1$  for $1d$, we first use a fixed dimension RG scheme, similar to the one used for the pure $1d$ KPZ equation~\cite{1dkpz-rg}. Setting $d=1$ in (\ref{flow1}) and (\ref{flow2}) above, we get
\begin{eqnarray}
 \frac{dg}{dl}&=&g\left[1-\frac{9g}{4}-\tilde g\right],\\
 \frac{d\tilde g}{dl}&=&\tilde g\left[1-\frac{3g}{2}-2\tilde g\right].
\end{eqnarray}
Setting $dg/dl=0=d\tilde g/dl$, apart from the Gaussian (trivial) fixed with ($g^*,\,\tilde g^*)=(0,0) $, that is globally unstable, we get three possible nontrivial fixed points:
\begin{enumerate}
 \item FP1: ($g^*,\,\tilde g^*)=(4/9,0) $, which is stable in the $g$-direction, but unstable in the $\tilde g$ direction.
 \item FP2: ($g^*,\,\tilde g^*)=(0,1/2)$, which is stable in the $\tilde g$-direction, but unstable in the $g$-direction.
 \item FP3: ($g^*,\,\tilde g^*)=(1/3,1/4)$, which is stable along both the $g$- and $\tilde g$-directions. This is the only globally stable fixed point.
 
 Eigenvalues of the stability matrix of the globally stable fixed points are -1 and -1/4.
 
 At the globally stable fixed point, we find (i) $z=4-g^*\frac{3}{4} + \tilde g^*=4-1/4+1/4=4$, and (ii) $\chi=1/2$. Thus, the scaling exponents are identical to their values in the linear theory. This we believe to be just fortuitous. Higher order corrections are likely to change this. 
\end{enumerate}
See Fig.~\ref{flow-short} for a schematic RG flow diagram in the $g-\tilde g$ plane.

Closed to $d=2$ but below it, we employ an $\epsilon$-expansion to the leading order in $\epsilon$, defined by $d=2-\epsilon,\,\epsilon>0$. We notice that (\ref{flow1}) becomes independent of $\tilde g$ to the leading order in $\epsilon$. Then by using (\ref{flow1}) and (\ref{flow2}) we 
\begin{equation}
 g^*=\frac{4\epsilon}{3},\,\tilde g^*=\frac{\epsilon}{3}.
\end{equation}
This is a globally stable fixed point. We do not mention the other fixed points, which are not globally stable. Surprisingly and related to the fact that the RG flow of $g$ is independent of $\tilde g$ to the leading order in $\epsilon$, the scaling exponents are {\em identical} to their values for the pure CKPZ equation:
\begin{equation}
 z=4-\frac{\epsilon}{3},\,\chi = \frac{1}{3}\epsilon.\label{expo-eps}
\end{equation}
We believe this is fortuitous. Higher order corrections are expected to make the scaling exponents to depend on $\tilde g^*$; this can already be seen from the fixed dimension RG results at $1d$. We note that the scaling exponents (\ref{expo-eps}) with $\epsilon=1$ do not reduce to their values obtained for $d=1$; we believe  { this is due to the limitation of the small-$\epsilon$ expansion used to obtain (\ref{expo-eps}}).

We now study the higher dimensional $d>2$ case. Setting $d=2-\epsilon,\,\epsilon<0$ (such that $d>2$), we get to the leading order in $\epsilon$
\begin{eqnarray}
 \frac{dg}{dl}&=& g\left[-\epsilon -\frac{3g}{4} \right],\label{flow3}\\
 \frac{d\tilde g}{dl}&=&\tilde g\left[-\epsilon -\tilde g-\frac{g}{2}\right].\label{flow4}
\end{eqnarray}
Flow equations (\ref{flow3}) and (\ref{flow4}) have only one physically acceptable solution: $g^*=0,\,\tilde g^*=0$, which is stable and which is identical to the Gaussian fixed point, confirming that $d=2$ is the upper critical dimension of the model, which is same as that for the pure CKPZ model. At this fixed point, unsurprisingly, the scaling exponents are identical to those in the linear theory.

\subsubsection{Long-ranged disorder}\label{long-dis}

To extract the universal scaling with long-ranged disorder, we follow the calculational scheme outlined for short-ranged disorder above. We start with action functional (\ref{action2}). The relevant one-loop Feynman diagrams are shown in Appendix~\ref{long-diag}.
 We use the same rescaling of the height field $h$ as given in (\ref{h-scale}). This procedure result into the following recursion relations
\begin{eqnarray}
\frac{d\nu}{dl}&=&\nu\left[z-4+ g\frac{4-d}{4d} + \tilde g \left(1+\frac{\alpha-2}{d}\right)\right],\\
 \frac{dD_h}{dl}&=&D_h\left[z-2-d-2\chi\right],\\
 \frac{d\lambda}{dl}&=& \lambda\left[z+\chi-4 + 2\tilde g - \frac{4}{d}\tilde g\right],\\
\frac{d}{dl}(\lambda_1^2D_F)&=&\lambda_1^2 D_F\left[2z-d+\alpha-6 -\frac{4\tilde g}{d}\right].
\end{eqnarray}
The  flow equations for $g$ and $\tilde g$ reads
\begin{eqnarray}
\frac{dg}{dl}&=&g\left[2-d+4\tilde g - \frac{8}{d}\tilde g - 3g \frac{4-d}{4d}- 3\tilde g\left(1+\frac{\alpha-d}{2}\right)\right],\nonumber\\ \label{flow3l}\\
\frac{d\tilde g}{dl}&=& \tilde g\left[2-d+ \alpha - \frac{4\tilde g}{d} - 2g \frac{4-d}{4d} - 2\tilde g \left(1+\frac{\alpha-2}{d}\right)\right].\label{flow4l}\nonumber\\
\end{eqnarray}
Clearly, the upper critical dimension of $\tilde g$ is $d=2+\alpha$, whereas that for $g$ is still 2.
We focus on the one-dimensional case, which is below the critical dimensions of both $g$ and $\tilde g$. We use  a fixed dimension RG, similar to the short-ranged case above. Apart from the Gaussian fixed point $(0,0)$, which is globally unstable, there are three more fixed points:\\\\
(i) FP1: $g^*=4/9,\,\tilde g^*=0$. Near this fixed point
\begin{equation}
 \frac{d\tilde g}{dl}=\tilde g\left[\alpha +\frac{1}{3}\right]>0
\end{equation}
for all $\alpha\geq 0$. Thus this fixed point is {\em always} globally unstable.\\\\
(ii) FP2: $g^*=0,\,\tilde g^* = 1/2$. Near this fixed point, writing $\tilde g=\tilde g^*+\delta\tilde g$,
\begin{equation}
 \frac{d\delta\tilde g}{dl}<0,
\end{equation}
and
\begin{equation}
 \frac{dg}{dl}=\frac{g}{2}\left[1-3\alpha\right] >(<)0
\end{equation}
if $\alpha<(>)1/3$. Thus, this fixed point is globally stable only if $\alpha>1/3$. The corresponding scaling exponents are given by
\begin{eqnarray}
 z&=&4-\frac{1}{2}(\alpha-1)>4,\\
 \chi&=&\frac{1}{2}\left[1-\frac{1}{2}(\alpha-1)\right]>1/2
\end{eqnarray}
so long as $\alpha<1$, which is what we consider here. Thus, the dynamics is slower and the surface is rougher than that in the linear theory.
\\\\
(iii) FP3: $g^*=(1+\alpha)(1-3\alpha)/3,\,\tilde g^*=(1+3\alpha)/4$. Clearly this fixed point ceases to exist for $\alpha>1/3$, since $g$ cannot be negative. For $0\leq \alpha\leq 1/3$, this fixed can exist. We now check the linear stability of this fixed point. The stability matrix $\cal M$ is given by
\begin{eqnarray}
{\cal M}=\left(\begin{array}{cc}
-\frac{3}{4}(1+\alpha)(1-3\alpha) & -\frac{(1+\alpha)(1+3\alpha)(1-3\alpha)}{3}\\
-\frac{3}{8}(1+3\alpha) & -\frac{(1+\alpha)(1+3\alpha)}{2}
\end{array}
\right).\end{eqnarray}
The eigenvalues $\tilde\Lambda$ of $\cal M$ are
\begin{eqnarray}
 &&\tilde \Lambda=-\frac{1}{8}\Big[(1+\alpha)(5-3\alpha)\nonumber \\&\pm&\sqrt{(1+\alpha)^2(5-3\alpha)^2-16(1+\alpha)(1-3\alpha)(1+3\alpha)}\; \Big].\nonumber\\
\end{eqnarray}
These are negative for $\alpha <1/3$, implying global stability of this fixed point.
Thus for $\alpha<1/3$, the globally stable fixed point is $g^*=(1+\alpha)(1-3\alpha)/3,\,\tilde g^*=(1+3\alpha)/4$. The associated scaling exponents are
\begin{eqnarray}
z&=&4+\alpha\\
\chi&=&\frac{1}{2}(1+\alpha).
\end{eqnarray}
Thus, both dynamic and the roughness exponents are {\em bigger} than their counterparts in the linear theory. Thus the surface is rougher and the dynamics is slower. Furthermore, these exponents reduce to their corresponding values for short-ranged disorder at $1d$ if we set $\alpha=0$.

\begin{figure}[htb]
\includegraphics[width=7cm]{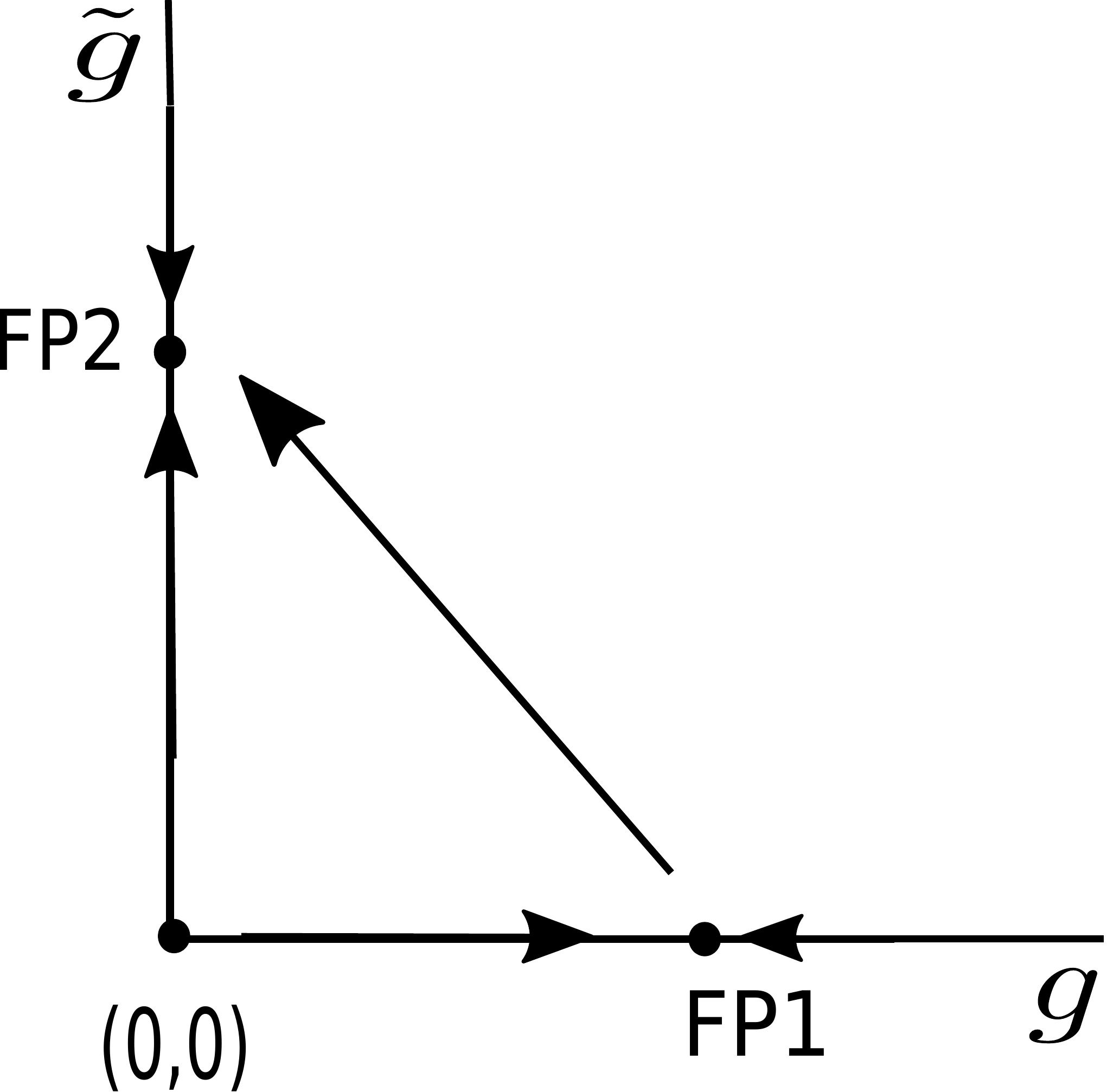}\\
\includegraphics[width=7cm]{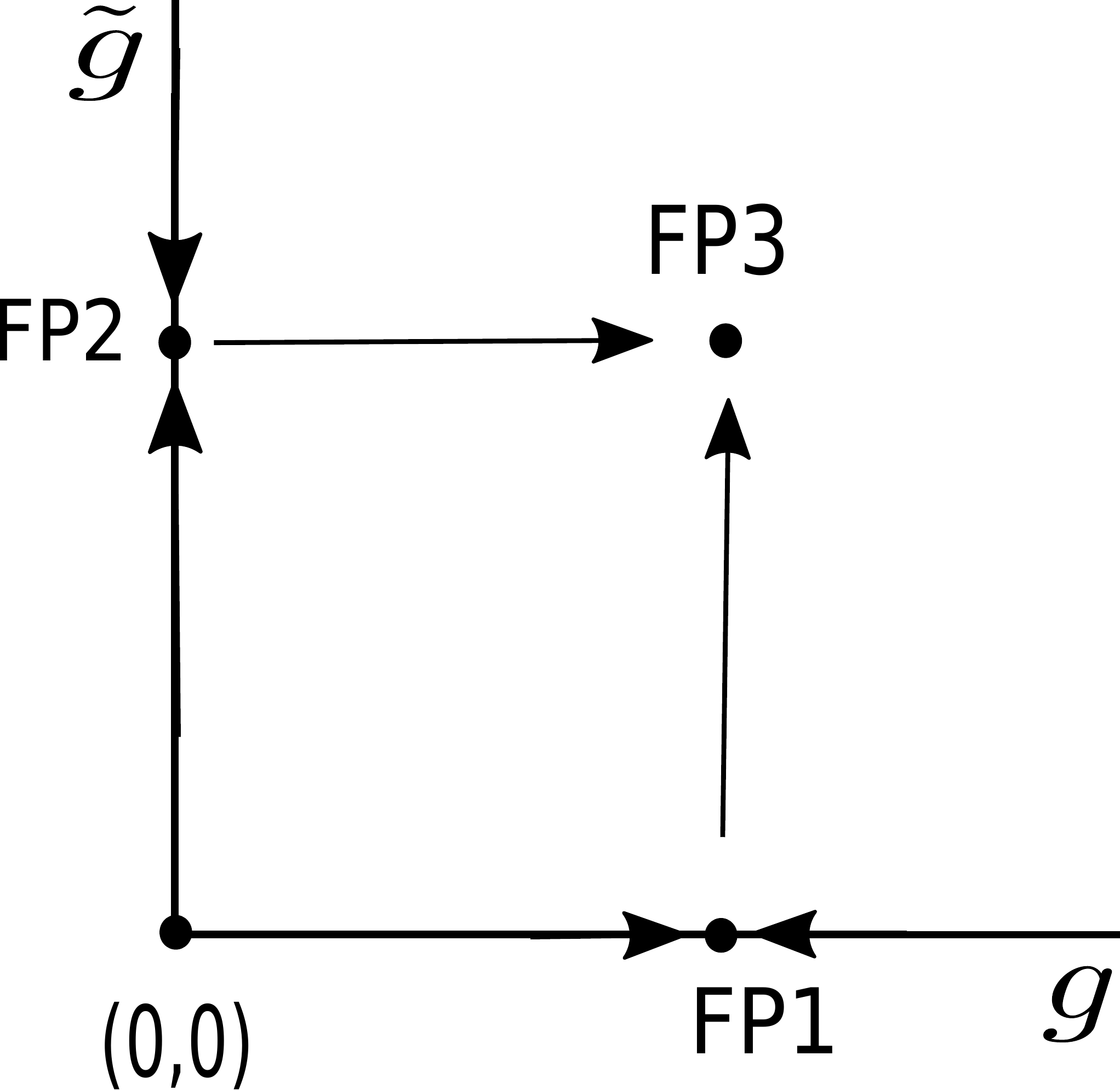}
\caption{RG flow diagrams in the $g-\tilde g$ plane. (top) $\alpha>1/3$, (bottom) $0<\alpha<1/3$ (this is topologically identical to the RG flow diagram for short-ranged disorder, as shown in Fig.~\ref{flow-short}). { Stable and unstable fixed points are shown: (top) FP2=(0,1/2) and (bottom) FP3=$(1+\alpha)(1-3\alpha)/3,\,(1+3\alpha)/4$ are the stable fixed points}; see text.}\label{rg-flow-long}
\end{figure}

The RG flow diagrams for $\alpha<1/3$ and $\alpha>1/3$ in the $g-\tilde g$ plane are shown in Fig.~\ref{rg-flow-long}.

We now briefly discuss the higher dimensional case. Noting that the upper critical dimension for $\tilde g$ is $d=2+\alpha$, we use an expansion in $\tilde\epsilon\equiv 2+\alpha-d$. The upper critical dimension of $g$ remains 2. Therefore, for sufficiently small $\tilde\epsilon$ with $\alpha>0$, we can neglect $g$. From (\ref{flow3}) we get
\begin{equation}
 \tilde g^*=\frac{\tilde\epsilon}{2}\frac{\alpha}{1+\alpha},
\end{equation}
which is a linearly stable fixed point.
This gives for the dynamic and the roughness exponents
\begin{eqnarray}
 z&=&4-\frac{\tilde\epsilon\alpha}{2(\alpha+1)},\\
 \chi&=&\frac{1}{2}\left[-\frac{\tilde\epsilon\alpha}{2(\alpha+1)}-\alpha+\tilde\epsilon\right].\label{long-chi1}
\end{eqnarray}
Interestingly, close to $d=2+\alpha$, i.e., for small $\tilde\epsilon$, $z$ is actually {\em smaller} than 4, its value in the linear theory, indicating that the disorder now makes the dynamics {\em faster}. { Furthermore, $\chi$ can be
negative if $\alpha$ is sufficiently large, i.e., when
$\tilde \epsilon < 2 \alpha (\alpha+1)/(\alpha+2)$).  A negative $\chi$ implies a smooth surface, e.g., the $\chi$ for the pure CKPZ equation or the 
KPZ equation (in its smooth phase) for $d>2$ are negative. Thus, by increasing $\alpha$, the disordered CKPZ model undergoes a rough-to-smooth transition for a sufficiently large $\alpha$. This opens up an intriguing possibility of smoothening by long-ranged disorder.  Further theoretical investigation would be useful to explore this feature.}

\section{Summary and outlook}\label{summ}

To summarise, we have studied the conserved KPZ equation that couples with quenched disorders. The coupling with the quenched disorder is such that is respects the symmetry of the pure CKPZ equation under a constant shift of the height field. Of course, it has no Galilean invariance, just as the pure {CKPZ equation} itself does not~\cite{janssen}. We have considered both the short-ranged and long-ranged quenched disorders. We find that with short-ranged disorder, the universal properties are affected by the disorder at $1d$, where as for higher $d\geq 2$, quenched { noise} is irrelevant (in a RG sense). With long-ranged disorder, quenched disorder continues to remain relevant at $2d$ and even higher, controlled by the spatial scaling of the variance of the Gaussian-distributed disorder, which is parametrised by $\alpha$ here. Notice that the values of the scaling exponents at $1d$ obtained by using a fixed dimension dynamic RG scheme quantitatively differ from what one would get for the exponents by setting $\epsilon=1$ or $\tilde\epsilon=1$. { We believe these mismatches are due to the limitations of the  small-$\epsilon$ expansion that we set up.} Higher order perturbative calculations and/or numerical simulations should be useful to extract quantitatively more accurate values of the scaling exponents. Nonetheless, we expect the general conclusions drawn here for the universality classes should be true.

 It will be instructive to consider and study microscopic lattice-gas models that belong to the same universality class as the continuum equation (\ref{ckpz-dis}). A possible route would be to suitably generalise the conserved restricted solid on solid (RSOS) model for surfaces (see, e.g., Ref.~\cite{ckpz}). In the usual $d$-dimensional conserved RSOS models, a site is randomly chosen and its height is increased by a unit and correspondingly reducing the height of neighbour by the same amount, while ensuring that the resulting configuration remains ``restricted'', i.e., the height difference between the neighbouring sites does not exceed one. Quenched disorder may be introduced by making the probability of the ``height exchange'' process bond-dependent in a quenched (or time-independent) manner. At a technical level, this may in-principle be achieved by making the probability that a bond is updated via the stochastic height exchange process quenched or time-independent with a distribution having spatially short- or long-ranged variances. { Yet another way for numerical verification of our results here would be to numerically integrate the continuum equation (\ref{ckpz-dis}) directly. This may be done, e.g., using the pseudospectral method~\cite{pseudo-kpz,abhik-erwin}, in $d$-dimensions. Although pseudospectral methods are known to produce numerical results with good accuracy. We however note with caution that applying it in the present context could be challenging because of the wide ranging time-scales involved among the different Fourier modes due to the fourth order diffusion operator in (\ref{ckpz-dis}).  These studies will be considered in the future.}

For simplicity, we have made a particular choice for coupling the disorder with the height field, such that the spatial average of the corresponding conserved current vanishes identically. Recently, the pure CKPZ equation has been generalised by inclusion of another nonlinear term that is as relevant as the existing nonlinear term of the pure CKPZ equation, but makes the current generally nonzero at zero wavevector~\cite{mike}. By using RG calculations, this new term has been shown to be a relevant perturbation on the pure CKPZ universality class. For instance, it introduces a roughening transition, absent in the CKPZ equation. It will be interesting to see how quenched disorders may affect this roughening transition. In fact, coupling with the quenched can also be generalised to make the disorder-dependent current to have a non-zero spatial average. Whether or not the roughening transition elucidated in Ref.~\cite{mike} survived such perturbation from the quenched disorder should form an interesting future study. 

Our work may be extended in a variety of ways. In this work, we have assumed the quenched disorder to couple with the height field multiplicatively. One could relax this and consider appropriate additive coupling with the quenched disorder, such that the height field dynamic will be subject to not just an additive annealed noise (as here), but also an additive quenched noise. Such additive quenched noise is found to be relevant and give rise to complex scaling behaviour for the ordinary KPZ equation~\cite{astik-prr}. It would be interesting to consider an analogous study for the quenched disordered CKPZ equation. Further, one may also study how the scaling behaviour of the CKPZ equation coupled with another dynamic field~\cite{tirtha} is affected by quenched disorder. We hope our work will provide impetus to further studies in these general directions.

\section{Acknowledgement}

The author thanks Abhik Basu (SINP, Kolkata) for helpful discussions and critical reading of the manuscript.

\appendix

\section{Renormalisation group calculations}

We discuss here in details the RG calculations. To that end, we first obtain the disorder-averaged action functional corresponding to the dynamical equation of motion (\ref{ckpz-dis}), and then apply one-loop perturbation theory on it.

\subsection{Disorder-averaged action functional}\label{short-action}

The RG calculations are greatly facilitated in terms of the generating functional for Eq.~(\ref{ckpz-dis}). We find
\begin{equation}
 {\cal Z}=\int {\cal D} h{\cal D}\hat h \exp (-S_F),
\end{equation}
where $\hat h ({\bf r},t)$ is the dynamic conjugate field of $h({\bf r},t)$~\cite{janssen}. The disorder-dependent action functional $S_F$ reads
\begin{widetext}
\begin{equation}
 S_F=\int d^dx dt D_h \hat h \nabla^2 \hat h + \int d^dx dt \hat h \left[\partial_t h + \nabla^2 \{\nu\nabla^2 h + \frac{\lambda}{2} ({\boldsymbol\nabla h})^2 + \lambda_1 {\bf F}\cdot {\boldsymbol\nabla}h\}\right].\label{action-inter}
\end{equation}
\end{widetext}
We now average over the the Gaussian-distributed quenched disorder $\bf F$. Only the term $\int d^dx dt \hat h \nabla^2\left[{\bf F}\cdot {\boldsymbol\nabla}h\right]=\int d^dx dt (\nabla^2\hat h) {\bf F}\cdot {\boldsymbol\nabla}h$ is to be affected by the disorder-averaging. We find
\begin{widetext}
\begin{equation}
 \langle \exp \left[\lambda_1 \int d^dx dt (\nabla^2 \hat h){\bf F}\cdot {\boldsymbol\nabla} h\right]\rangle = \exp \left[\frac{\lambda_1^2}{2} \int d^dx_1 d^dx_2 \int dt_1 dt_2 \left[(\nabla^2 \hat h)\nabla_m h\right]_{{\bf x}_1,t_1} \langle F_m({\bf x}_1) F_n({\bf x}_2)\rangle \left[(\nabla^2 \hat h)\nabla_n h\right]_{{\bf x}_2,t_2}\right].\label{gen-act}
\end{equation}

We now separately deal with short and long-ranged disorders. For short-ranged disorders, we use (\ref{short-vari}) to arrive at a disorder-averaged action

\begin{equation}
 {\cal S}\equiv \int d^dx dt D_h \hat h \nabla^2 \hat h + \int d^dx dt \hat h \left[\partial_t h + \nabla^2 \{\nu\nabla^2 h + \frac{\lambda}{2} ({\boldsymbol\nabla h})^2\right] -\lambda_1^2 D_F \int d^dx \int dt_1 dt_2 \left[(\nabla^2 \hat h)\nabla_m h\right]_{{\bf x},t_1}\left[(\nabla^2 \hat h)\nabla_m h\right]_{{\bf x},t_2}.\label{action1}
\end{equation}

For long ranged disorders, we use (\ref{long-vari}) to arrive at

\begin{eqnarray}
 {\cal S}&\equiv& \int d^dx dt D_h \hat h \nabla^2 \hat h + \int d^dx dt \hat h \left[\partial_t h + \nabla^2 \{\nu\nabla^2 h + \frac{\lambda}{2} ({\boldsymbol\nabla h})^2\right] -\lambda_1^2 D_F \int d^dx_1d^dx_2 \int dt_1 dt_2 \left[(\nabla^2 \hat h)\nabla_m h\right]_{{\bf x}_1,t_1}\nonumber \\ &\times&|{\bf x}_1 - {\bf x}_2|^{\alpha-d}\left[(\nabla^2 \hat h)\nabla_m h\right]_{{\bf x}_{ 2},t_2}.\label{action2}
\end{eqnarray}


It is convenient to write the time-nonlocal terms in (\ref{action1}) and (\ref{action2}) in the Fourier space. We get
 \begin{eqnarray}
  &&\int d^dx \int dt_1 dt_2 \left[(\nabla^2 \hat h)\nabla_m h\right]_{{\bf x},t_1}\left[(\nabla^2 \hat h)\nabla_m h\right]_{{\bf x},t_2} \nonumber \\&=& \int d^dkd^dq_1 d^dq_2\int d\omega d\Omega_1d\Omega_2 \left[q_1^2 \hat h({\bf q}_1, \Omega_1) i ({\bf k-q}_1)_m h({\bf k-q}_1,\omega -\Omega_1)\right]\nonumber \\&&\times\delta(\omega) \left[q_2^2 \hat h({\bf q}_2, \Omega_2) i ({\bf -k-q}_2)_m h({\bf -k-q}_2,-\omega -\Omega_2)\right],
 \end{eqnarray}
and

 \begin{eqnarray}
  &&\int d^dx_1d^dx_2 \int dt_1 dt_2 \left[(\nabla^2 \hat h)\nabla_m h\right]_{{\bf x}_1,t_1}|{\bf x}_1 - {\bf x}_2|^{\alpha-d}\left[(\nabla^2 \hat h)\nabla_m h\right]_{{\bf x}_2,t_2} \nonumber \\&&= \int d^dkd^dq_1 d^dq_2\int d\omega d\Omega_1d\Omega_2 \left[q_1^2 \hat h({\bf q}_1, \Omega_1) i ({\bf k-q}_1)_m h({\bf k-q}_1,\omega -\Omega_1)\right]\nonumber \\&&\times\delta(\omega)|k|^{-\alpha} \left[q_2^2 \hat h({\bf q}_2, \Omega_2) i ({\bf - k-q}_2)_m h({\bf -k-q}_2,-\omega -\Omega_2)\right].
 \end{eqnarray}

\end{widetext}

\subsection{Feynman diagrams for short-ranged disorder}\label{short-appen}

If we ignore the anharmonic terms in (\ref{action1}) or (\ref{action2}) then we can evaluate all the two-point correlation functions exactly. These in the Fourier space read
\begin{eqnarray}
 \langle \hat h({\bf k},\omega)\hat h({\bf -k},-\omega)\rangle &=&0,\\
 \langle \hat h ({\bf k},\omega)h(-{\bf k},-\omega)\rangle&=& \frac{1}{i\omega + \nu k^4},\\
 \langle \hat h (-{\bf k},-\omega)h({\bf k},\omega)\rangle&=& \frac{1}{-i\omega + \nu k^4},\\
 \langle h({\bf k},\omega)h({\bf -k},-\omega)\rangle &=& \frac{2D_hk^2}{\omega^2 +\nu^2k^4}.
 \end{eqnarray}

\begin{figure}[htb]
 \includegraphics[width=6cm]{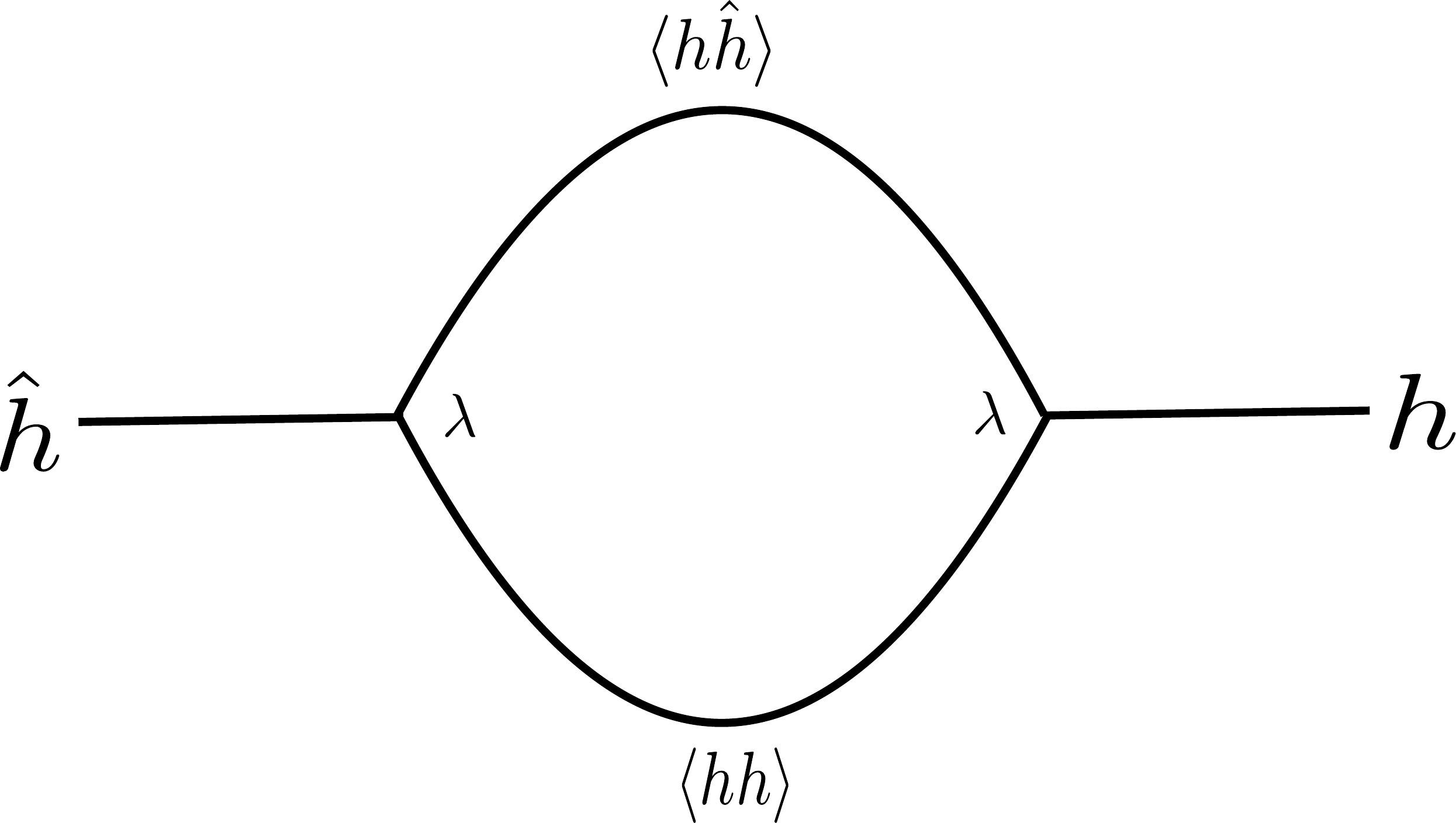}\\
 \includegraphics[width=6cm]{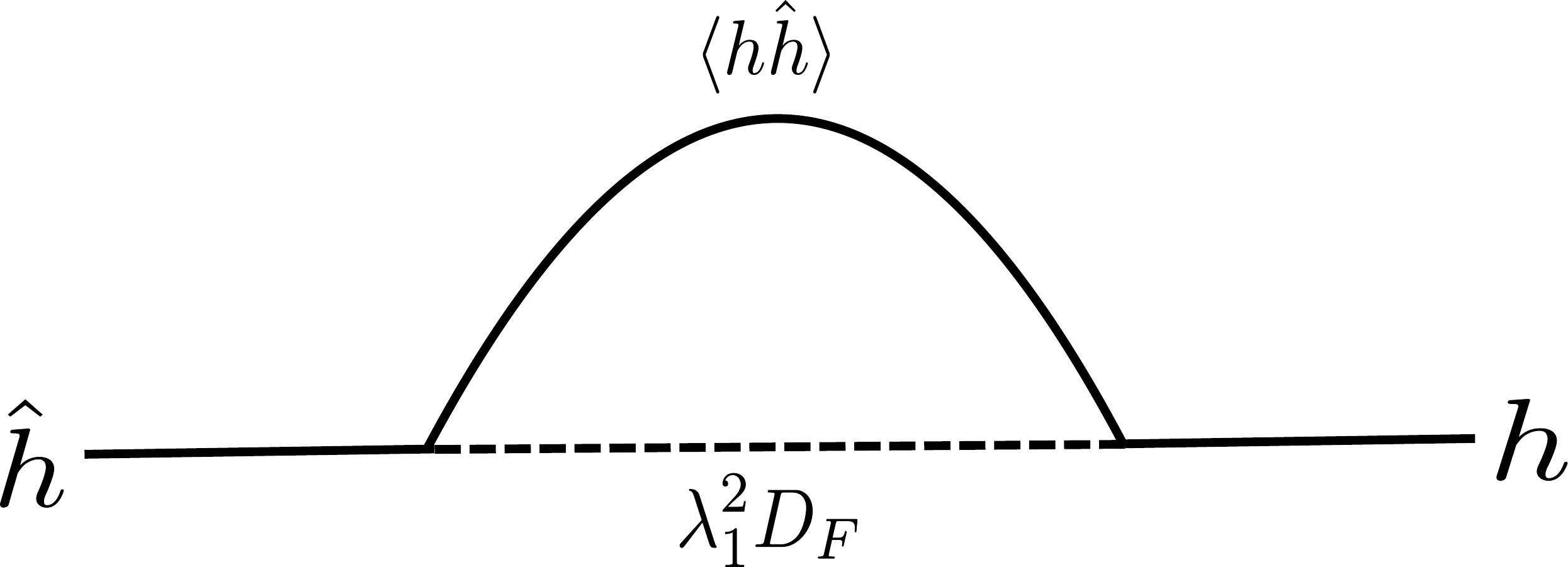}
 \caption{One-loop Feynman diagrams for $\nu$ with short-ranged disorder: (top) one-loop diagram that exists in the pure CKPZ problem, (bottom) one-loop diagram that originates from the short-ranged disorder (see text).}\label{nu-short-diag}
\end{figure}
\begin{figure}[htb]
 \includegraphics[width=6cm]{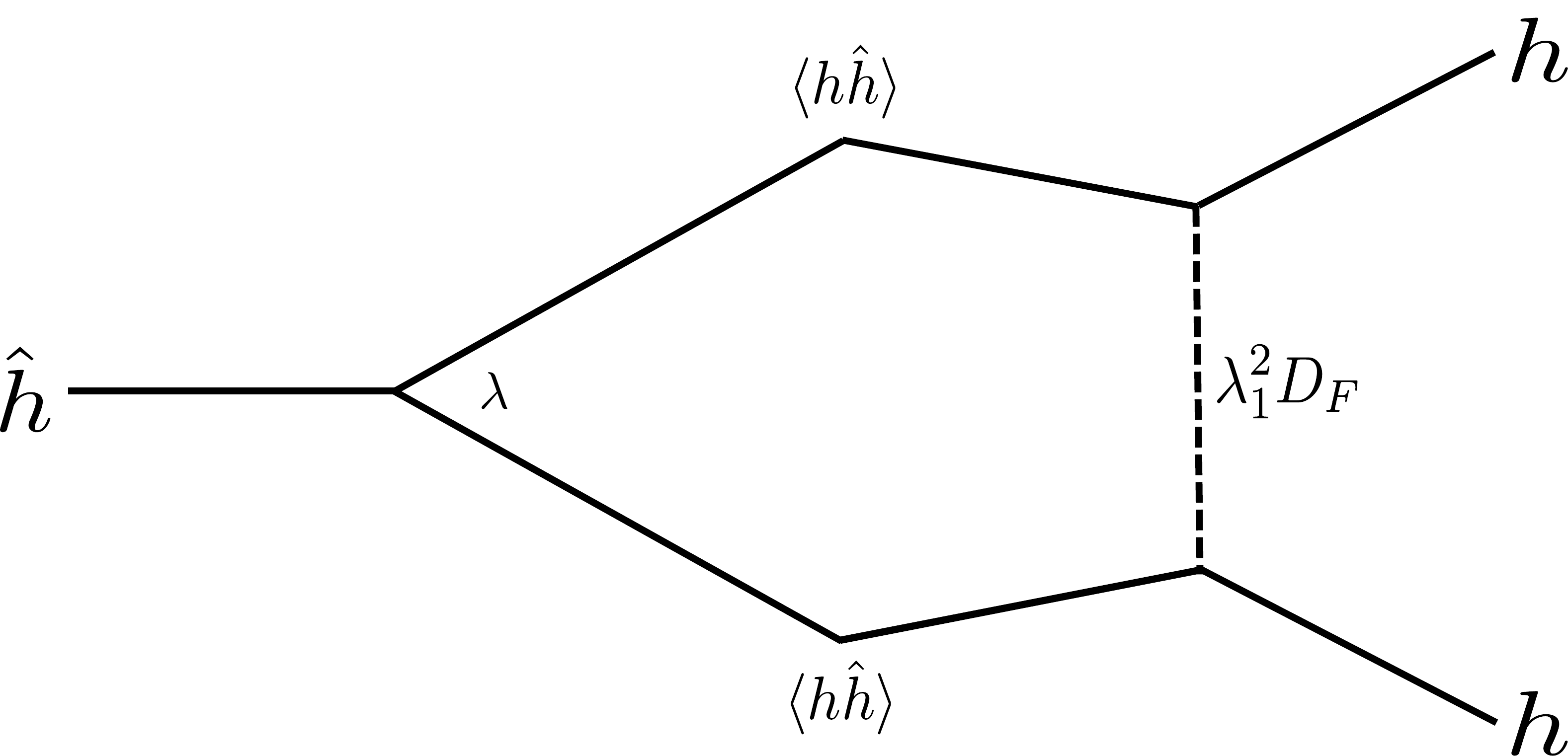}\\
 \includegraphics[width=6cm]{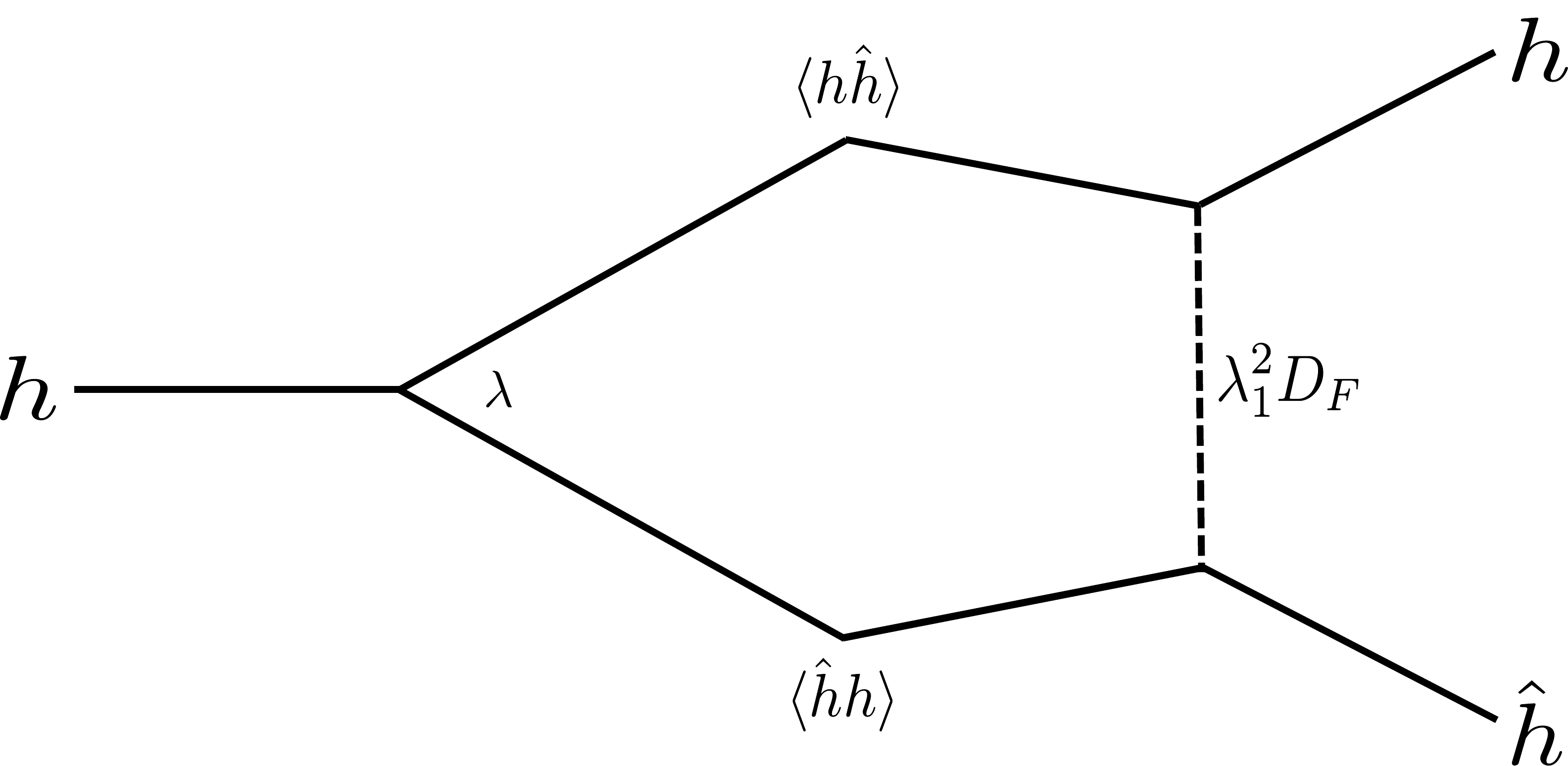}
 \caption{One-loop diagrams for $\lambda$ with short-ranged disorder. All these diagrams vanish for the pure CKPZ problem.}\label{lam-short-diag}
\end{figure}
\begin{figure}[htb]
 \includegraphics[width=6cm]{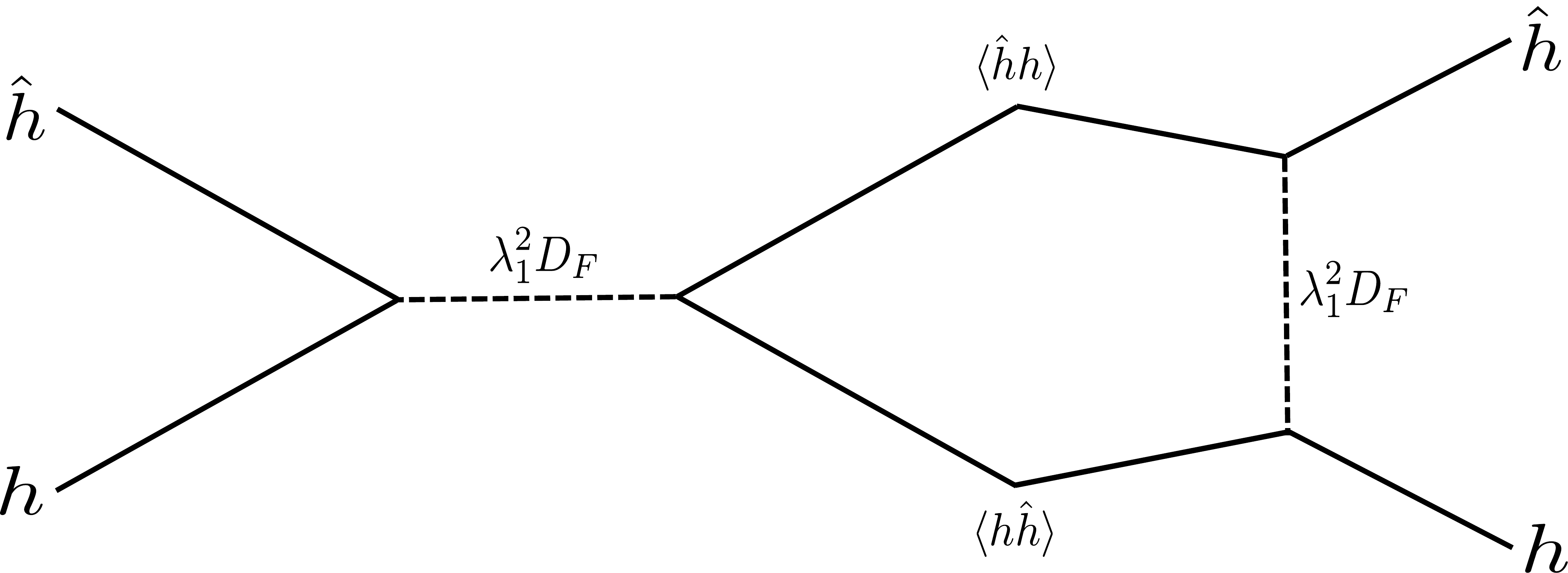}\\
 \includegraphics[width=6cm]{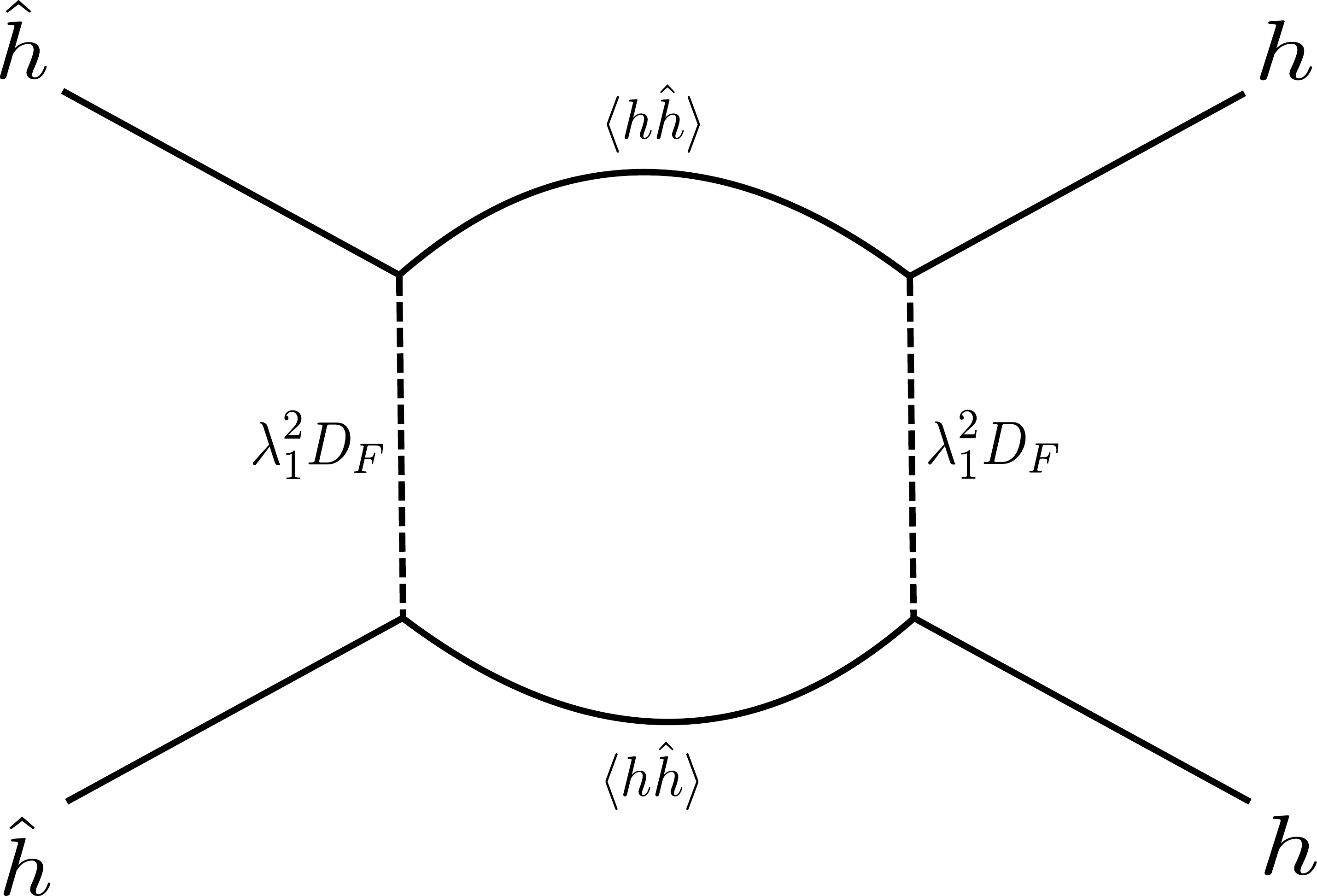}\\
 \includegraphics[width=6cm]{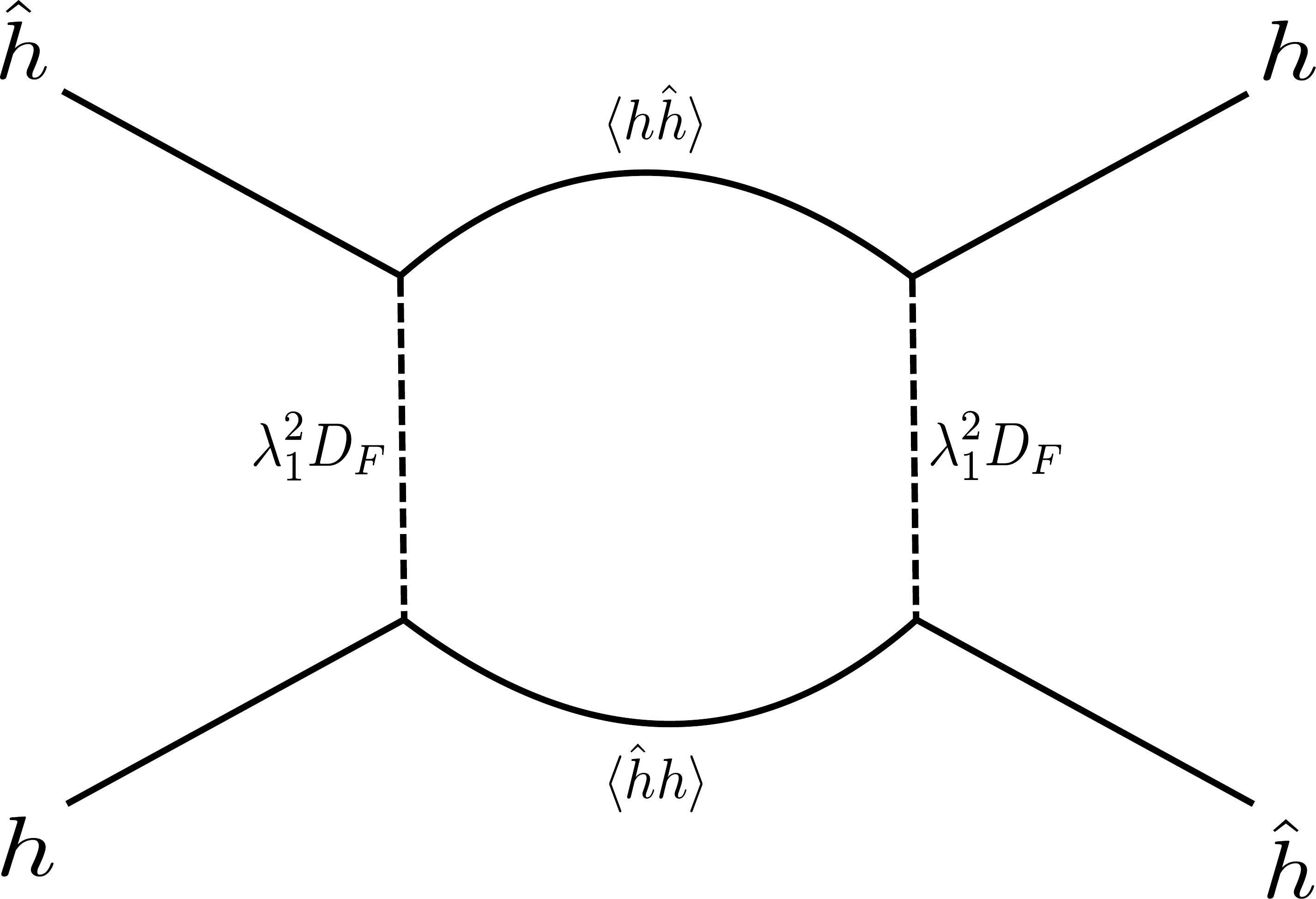}
 \caption{One-loop diagrams for $\lambda$ with short-ranged disorder. All these diagrams vanish for the pure CKPZ problem.}\label{lam1-short-diag}
\end{figure}

We now give here the one-loop Feynman diagrams for $\nu,\,\lambda$ and $\lambda_1^2D_F$ with short-ranged disorder in Fig.~\ref{nu-short-diag}, Fig.~\ref{lam-short-diag} and Fig.~\ref{lam1-short-diag}, respectively. In each of the diagrams, a broken line represents the short-ranged disorder.

There are three more infra-red divergent diagrams for $\lambda_1^2D_F$, over and above those shown in Fig.~\ref{lam1-short-diag}, the total correction for $\lambda_1^2D_F$ coming from
such three diagrams is zero. 

Combining all the fluctuation corrections, we obtain the fluctuation-corrected $\nu$, $\lambda$ and $\lambda_1^2 D_F$. We get

\begin{widetext}
\begin{eqnarray}
 \nu^<&=& \nu \left[1+\frac{\lambda^2 D_h}{\nu^3}k_d \frac{4-d}{4d}\int_{\Lambda/b}^\Lambda dq \,q^{d-3} + \frac{\lambda_1^2 D_F}{\nu^2} \left(1-\frac{2}{d}\right)k_d\int_{\Lambda/b}^\Lambda dq\,q^{d-3}\right],\\
 \lambda^<&=& \lambda\left[1+2\frac{\lambda_1^2D_F}{\nu^2}k_d\int_{\Lambda/b}^\Lambda dq\,q^{d-3} -\frac{4}{d}\frac{\lambda_1^2D_F} {\nu^2}k_d\int_{\Lambda/b}^\Lambda dq\,q^{d-3}\right],\\
 \left(\lambda_1^2 D_F\right)^< &=& \lambda_1^2D_F\left[1-\frac{6\lambda_1^2D_F}{\nu^2d}k_d \int_{\Lambda/b}^\Lambda dq\, q^{d-3} + 2\frac{\lambda_1^2D_F}{\nu^2}k_d \int_{\Lambda/b}^\Lambda dq\, q^{d-3}\right] \label{a14}
\end{eqnarray}
\end{widetext}
for short-ranged disorder. In the differential recursion relations that follow, we set $\Lambda=1$.

\subsection{Feynman diagrams for long-ranged disorder}\label{long-diag}

\begin{figure}
 \includegraphics[width=6cm]{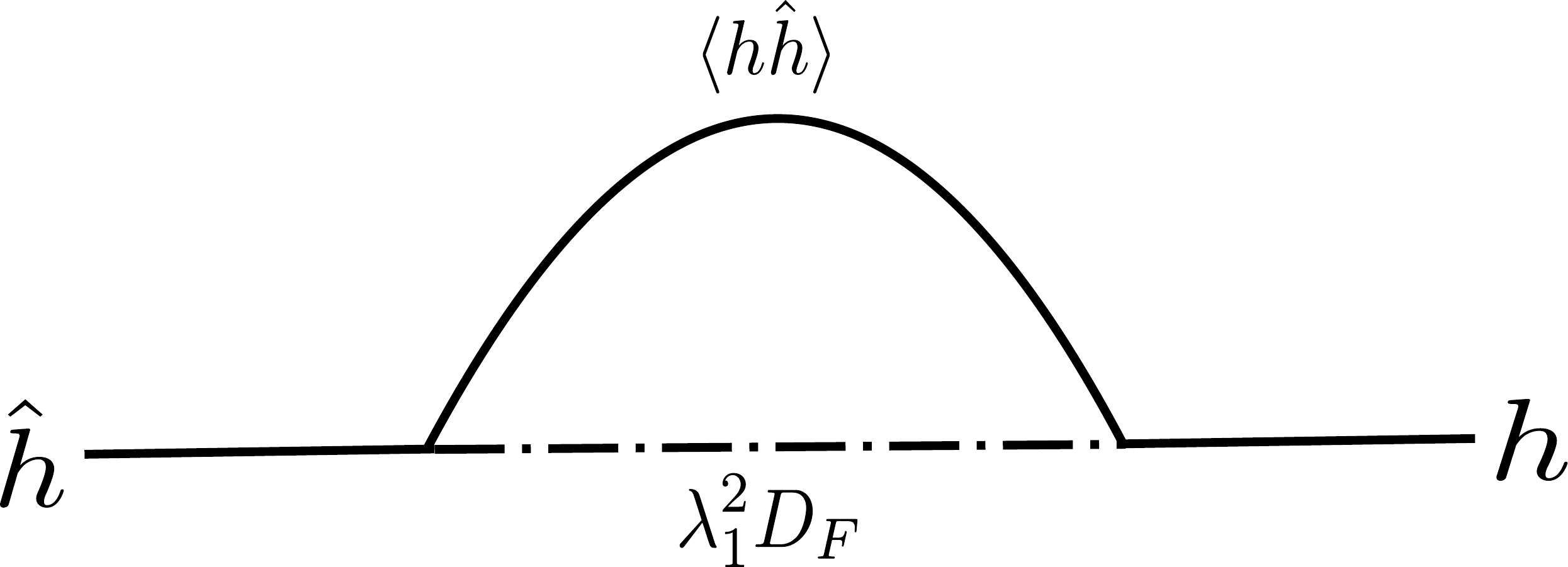}
 \caption{One-loop diagram for $\nu$ with long-ranged disorder.}\label{nu-long}
\end{figure}

\begin{figure}
 \includegraphics[width=6cm]{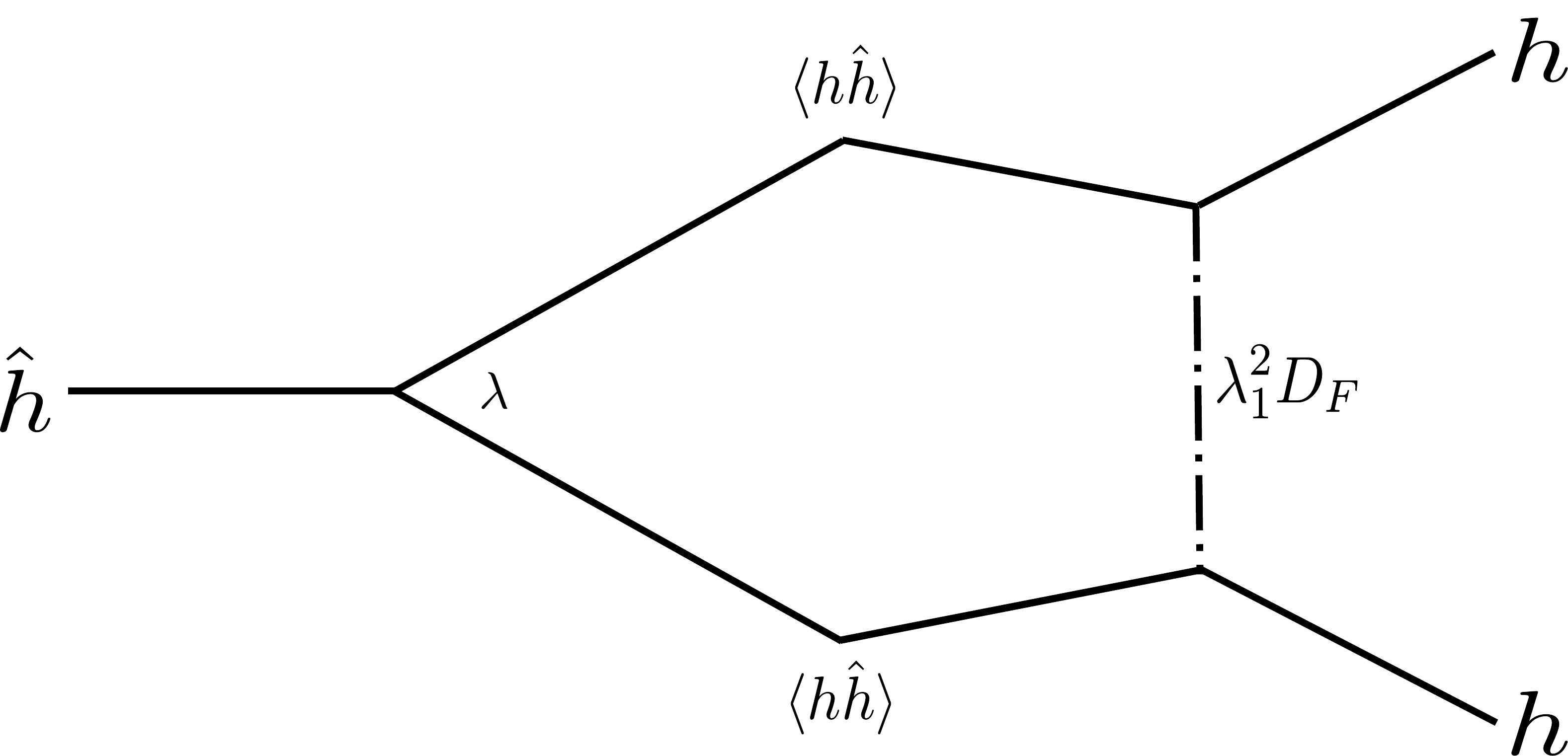}\\
  \includegraphics[width=6cm]{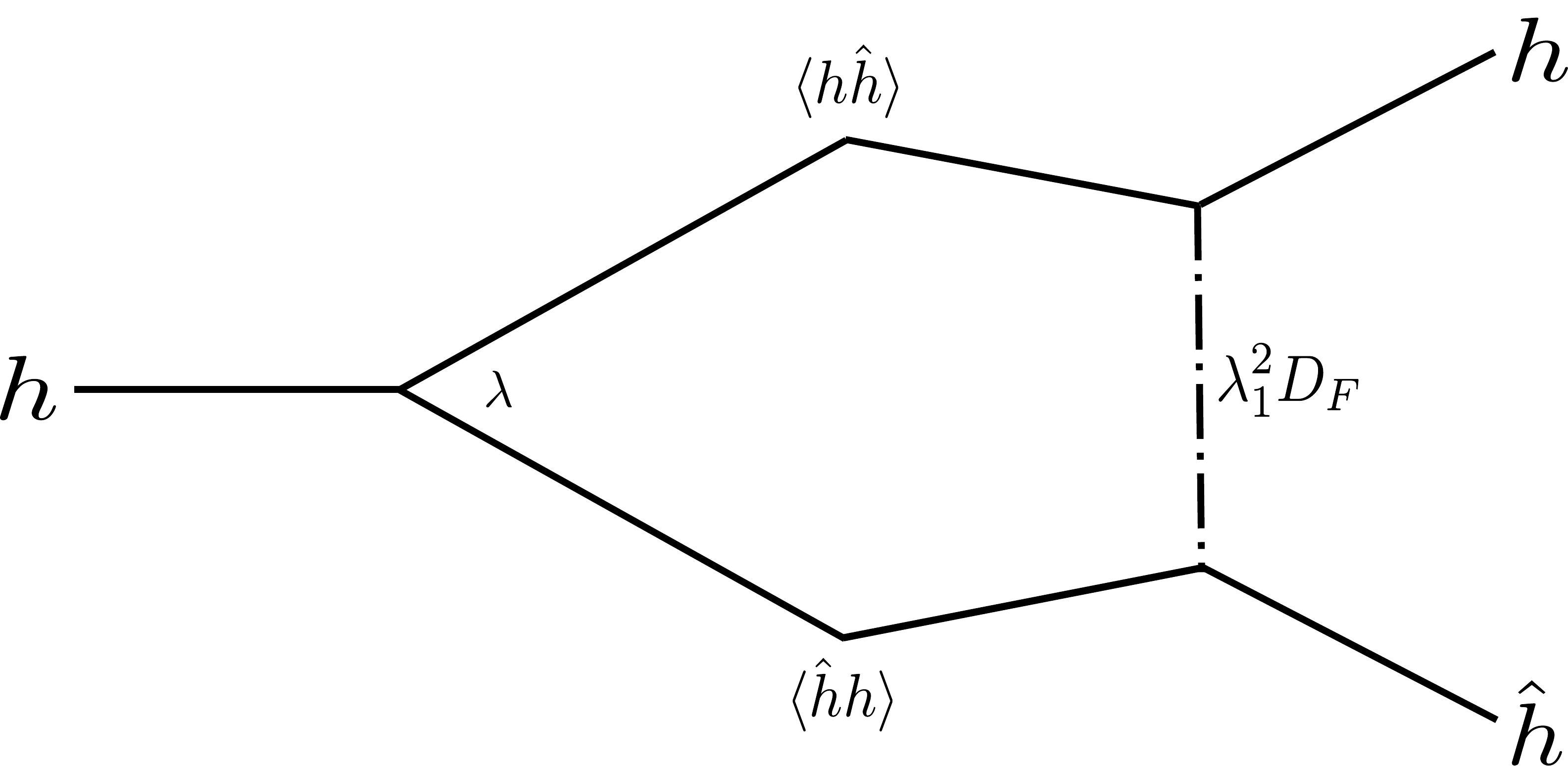}
 \caption{One-loop diagrams for $\lambda$ with long-ranged disorder.}\label{lam-long}
\end{figure}

\begin{figure}
 \includegraphics[width=6cm]{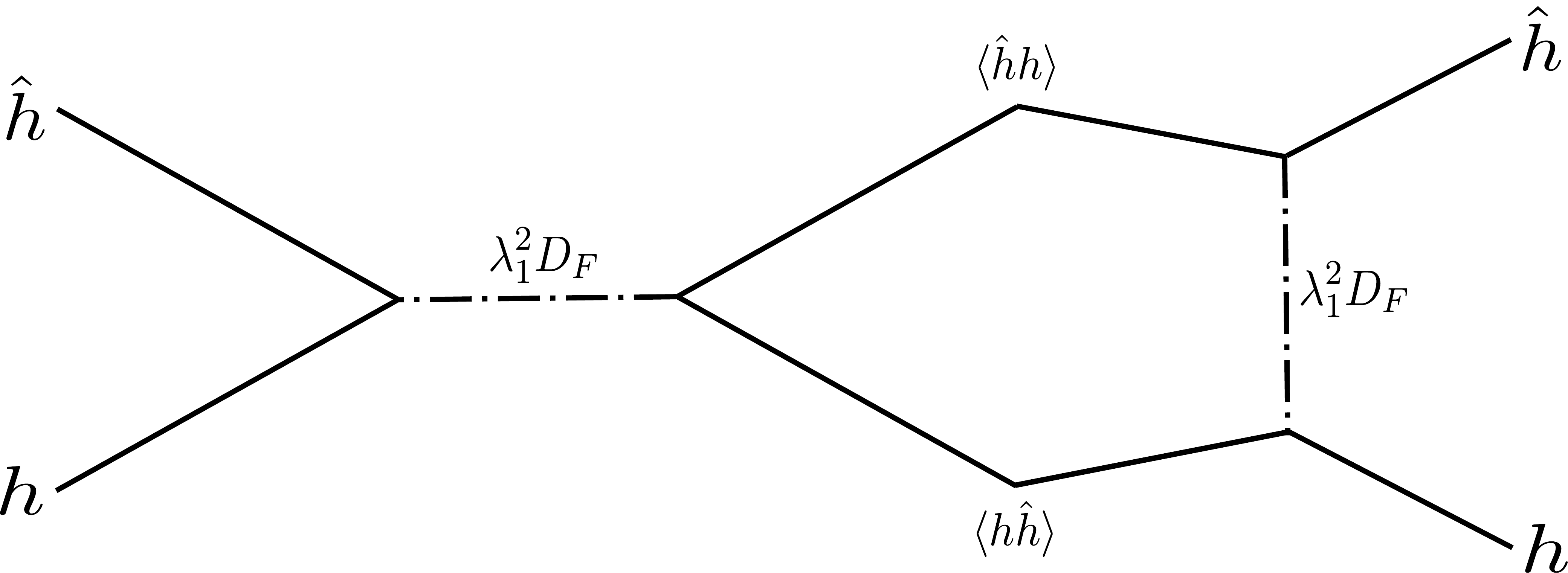}
 \caption{One-loop diagram for $\lambda_1^2D_F$ with long-ranged disorder.}\label{dis-long}
\end{figure}

We now give the one-loop Feynman diagrams for $\nu,\lambda$ and $\lambda_1^2D_F$ with long-ranged disorder in Fig.~\ref{nu-long}, Fig.~\ref{lam-long} and Fig.~\ref{dis-long}, respectively. In each of these diagrams, a dotted-dashed line represents the long-ranged disorder vertex. We have only shown the most relevant diagrams, all of which arise from the disorder-vertex. { The relevant diagrams with long-ranged disorder are topologically identical to some of the diagrams in the short-ranged disorder case. Nonetheless, we present them here separately to highlight the fact that the disorder lines (dot-dashed lines) in the Feynman diagrams for the long-ranged disorder case carry a factor $k^{-\alpha}$, where $k$ is a wavevector, whereas the disorder lines (dashed lines) in the Feynman diagrams  for the short-ranged disorder case carry only a constant factor.}

For long-ranged disorder ($\alpha>0$), the effective, fluctuation-corrected model parameters are
\begin{widetext}
 \begin{eqnarray}
  \nu^<&=&\nu \left[1+\frac{\lambda^2 D_h}{\nu^3}k_d \frac{4-d}{4d}\int_{\Lambda/b}^\Lambda dq \,q^{d-3} + \frac{\lambda_1^2 D_F}{\nu^2}\left(1+\frac{\alpha-2}{d}\right)k_d\int_{\Lambda/b}^\Lambda dq q^{d-3-\alpha}\right],\\
  \lambda^<&=& \lambda\left[1+2 \frac{\lambda_1^2D_F}{\nu^2}k_d \int_{\Lambda/b}^\Lambda dq q^{d-3-\alpha}-\frac{4}{d}\frac{\lambda_1^2D_F} {\nu^2}k_d\int_{\Lambda/b}^\Lambda dq\,q^{d-3-\alpha}\right],\\
  \left(\lambda_1^2D_F\right)^<&=& \lambda_1^2D_F\left[1-\frac{4\lambda_1^2D_F}{\nu^2 d}k_d\int_{\Lambda/b}^\Lambda dq q^{d-3-\alpha}\right].\label{a17}
 \end{eqnarray}

\end{widetext}
 { Notice that the one-loop diagrams contributing to $\lambda_1^2 D_F$ is more for short-ranged disorder (Fig.~\ref{lam1-short-diag}) than those for long-ranged disorder (Fig.~\ref{dis-long}). This is because the extra diagrams in Fig.~\ref{lam1-short-diag}, which exist even for the long-ranged case in principle, actually imply generation of short-ranged disorder in the long-ranged disorder case. Since for $\alpha>0$ the short-ranged disorder is irrelevant (in the RG sense) in the presence of the long-ranged disorder, we ignore these contributions for the long-ranged disorder case~\cite{wein}. As a result, setting  $\alpha=0$ does not reduce (\ref{a17}) to (\ref{a14}). }

\subsection{Rescaling}~\label{resc}

We discuss the rescaling of space, time and the long wavelength parts of the fields $h$ and $\hat h$. We lay our the details for the long-ranged disorder case; the corresponding rescaling for the short-ranged case can be easily retrieved by setting $\alpha=0$. We scale space and time as follows:
\begin{equation}
 {\bf x}\rightarrow b{\bf x},\;t\rightarrow b^zt,
\end{equation}
where $z$ is the dynamic exponent. Under these spatio-temporal rescaling, we let the long wavelength parts of $h$ and $\hat h$ to scale as
\begin{equation}
 h\rightarrow b^\chi h,\;\hat h \rightarrow b^{ \chi'} \hat h.
\end{equation}
Here, the rescaling factor $b>1$. We can now calculate the rescaling factors of the different terms in (\ref{action2}). For instance,
\begin{widetext}
\begin{enumerate}
 \item We get $\int d^dx dt \hat h \partial_t h \rightarrow b^{\chi+\chi'-d}\int d^dx dt\, \hat h^\prime \partial_{t}h^\prime.$
 Demanding that the coefficient of $\int d^dx dt \hat h \partial_t h $ remains unity after rescaling, we get $\chi+\chi'=-d$.
 \item Next, $\int d^dx d^dx'\int dt\,dt'\lambda_1^2 D_F\left(\nabla^2\hat h\nabla_m h\right)_{{\bf x},t} \tilde D_h(|{\bf x-x'}|)\left(\nabla^2\hat h\nabla_m h\right)_{{\bf x'},t'}\rightarrow \\b^{2z-d+\alpha-6}\int d^dx d^dx' dt dt' \lambda_1^2 D_F \left(\nabla^2\hat h\nabla_m h\right)_{{\bf x},t} \tilde D_h(|{\bf x-x'}|)\left(\nabla^2\hat h\nabla_m h\right)_{{\bf x'},t'}.$ This gives $(\lambda_1^2D_F)\rightarrow b^{2z-d+\alpha-6}(\lambda_1^2D_F)$.
 \item Then, $\lambda \int d^dx dt (\nabla^2\hat h) ({\boldsymbol\nabla}h)^2\rightarrow b^{-z+2\chi+\chi'-d-4}\lambda \int d^dx dt (\nabla^2\hat h) ({\boldsymbol\nabla}h)^2$. This gives $\lambda\rightarrow b^{\chi+z-4}\lambda$.
 \item Next, $\int d^dx dt D_h \hat h \nabla^2 \hat h \rightarrow b^{2{ \chi'} +z+d-2}\int d^dx dt D_h \hat h \nabla^2 \hat h$. This gives $D_h\rightarrow b^{z-d-2-2\chi}D_h$.
 \item Lastly $\int d^dx dt \nu\hat h\nabla^4 h \rightarrow b^{\chi+\chi'+z+d-4}\int d^dx dt \nu\hat h\nabla^4 h$. This gives $\nu\rightarrow b^{z-4}\nu$.
\end{enumerate}

\end{widetext}
These rescaling of the model parameters together with the one-loop fluctuation corrections lead the different flow equations in the main text.

\end{document}